\documentclass[runningheads]{llncs}
\usepackage[T1]{fontenc}
\usepackage{graphicx}
\usepackage[colorlinks,
            linkcolor=blue,
            anchorcolor=blue,
            citecolor=blue]{hyperref}
\usepackage{hyperref}
\usepackage{booktabs} 
\usepackage{pifont} 
\usepackage{graphicx}
\usepackage{subcaption}
\usepackage[T1]{fontenc}
\usepackage{graphicx}
\usepackage[misc]{ifsym}
\usepackage{amsfonts}
\usepackage{amssymb}
\usepackage{amsmath}

\usepackage{mathrsfs}
\usepackage{indentfirst}
\usepackage{multirow}
\usepackage{bm}
\usepackage{color,xcolor}
\usepackage{footnote}
\usepackage{float}
\usepackage{lipsum}
\usepackage[marginal]{footmisc}

\begin{document}

\title{Multimodal Cross-Task Interaction for Survival Analysis in Whole Slide Pathological Images}

\author{
    Songhan Jiang \inst{1} \and
    Zhengyu Gan \inst{1} \and
    Linghan Cai\inst{1} \and
    Yifeng Wang\inst{2} \and \\  
     Yongbing Zhang\inst{1}\textsuperscript{(\Letter)}
}
\authorrunning{S. Jiang et al.}
\institute{
    School of Computer Science and Technology, Harbin Institute of Technology (Shenzhen), Shenzhen 518055, China \and
    School of Science, Harbin Institute of Technology (Shenzhen), Shenzhen 518055, China \\
    \email{\href{ybzhang08@hit.edu.cn}{ybzhang08@hit.edu.cn} }
}

\maketitle              

\footnote{S. Jiang, Z. Gan, and L. Cai—Contributed equally to this work.\\}

\begin{abstract}

Survival prediction, utilizing pathological images and genomic profiles, is increasingly important in cancer analysis and prognosis. Despite significant progress, precise survival analysis still faces two main challenges: 
(1) The massive pixels contained in whole slide images (WSIs) complicate the process of pathological images, making it difficult to generate an effective representation of the tumor microenvironment (TME).
(2) Existing multimodal methods often rely on alignment strategies to integrate complementary information, which may lead to information loss due to the inherent heterogeneity between pathology and genes. 
In this paper, we propose a Multimodal Cross-Task Interaction (MCTI) framework to explore the intrinsic correlations between subtype classification and survival analysis tasks. Specifically, to capture TME-related features in WSIs, we leverage the subtype classification task to mine tumor regions. Simultaneously, multi-head attention mechanisms are applied in genomic feature extraction, adaptively performing genes grouping to obtain task-related genomic embedding. With the joint representation of pathological images and genomic data, we further introduce a Transport-Guided Attention (TGA) module that uses optimal transport theory to model the correlation between subtype classification and survival analysis tasks, effectively transferring potential information. Extensive experiments demonstrate the superiority of our approaches, with MCTI outperforming state-of-the-art frameworks on three public benchmarks. \href{https://github.com/jsh0792/MCTI}{https://github.com/jsh0792/MCTI}.

\end{abstract}
%

\subsubsection{Acknowledgements}
This research received support by the National Natural Science Foundation of China (62031023 \& 62331011), the Shenzhen Science and Technology Project (JCYJ20200109142808034 \& GXWD20220818170353009), the Fundamental Research Funds for the Central Universities (Grant No. HIT. OCEF. 2023050), and the Beijing Hospitals Authority’Ascent Plan (DFL20190701).

\section{Introduction}
Survival analysis is a crucial topic in clinical prognosis research, aiming to predict the time elapsed from a known origin to events of interest, such as death and disease recurrence \cite{aalen2008survival,dey2022survival,nagy2021pancancer,shmatko2022artificial}. Accurate survival prediction is significant for clinical management and decision-making, benefiting patients by enabling healthcare professionals to tailor personalized treatment plans. Traditionally, survival analysis is time-consuming and labor-intensive to make a predictive diagnosis by pathologists~\cite{collins2015new,jackson2020single}. With the development of deep learning, survival analysis based on whole slide images (WSIs) and genomic profiles~\cite{MCAT,Porpoise,M3IF,CMTA} has shown massive potential for facilitating disease progression and treatment. 

Given the gigapixel resolution of WSIs (e.g., $40,000\times40,000$ pixels), the pathological image analysis is often formulated as a weakly supervised task using multiple instance learning (MIL). A WSI is developed as a bag containing multiple instances (patches) within the MIL framework. Existing MIL approaches \cite{ABMIL,DSMIL,CLAM} generally employ a two-stage architecture: initially using a deep neural network to extract instance features and subsequently aggregating them through a pooling function to obtain a bag representation utilized in downstream tasks. In the task of survival analysis, to better extract pathological features, WSISA~\cite{WIASSA} adopts k-means clustering in the MIL framework to capture representative patches. DeepAttnMISL~\cite{DeepAttnMISL} further introduces an attention pooling \cite{ABMIL} that adaptively aggregates the selected instances for improving bag representation. However, these methods could not effectively extract the tumor microenvironment (TME) contained in WSIs, as they ignore areas that may have critical information, like tumor cells and lymphocyte infiltration~\cite{dey2022efficient}, which are highly relevant to survival analysis. On the other hand, the subtype classification task requires the network to capture the tumor regions involved in WSI. Consequently, introducing a subtype classification task could potentially enrich the TME-related features, promoting survival analysis performance.

Genes expression corresponds to some morphological characteristics of pathological TME~\cite{medema2011microenvironmental,ramaswamy2001multiclass}, which is crucial for improving survival analysis. Most related works focus on solving the alignment problem among different modalities~\cite{9186053,MCAT,MoCAT,M3IF}. Pathomic Fusion~\cite{9186053} develops a tensor fusion module to fuse pathological and genomic features.
MCAT~\cite{MCAT} uses a multimodal co-attention module to identify instances from pathological images using genomic features as queries.
MoCAT~\cite{MoCAT} builds interactions between pathology and genomics through optimal transport, aligning genomic representations to pathological features. However, the alignment process inevitably loses modality-specific information and semantic differences between pathological images and genomic profiles. 
Unlike these methods, this paper explores the task correlation between subtype classification and survival analysis, optimizing the joint representation of genes and pathology through task interaction, aiming to enhance the effectiveness of survival analysis.

This paper proposes a Multimodal Cross-Task Interaction (MCTI) framework that integrates the subtype classification task for improving survival analysis. 
Specifically, MCTI leverages the tumor localization ability of attention-based multiple instance learning framework in the context of subtype classification task, effectively enriching TME-related features.
Meanwhile, multi-head attention mechanisms are used to process genes for adaptively grouping and embedding. With the joint representation of pathological image and genes, we perform a Transport-Guided Attention (TGA) based encoder-decoder for feature reconstruction, where TGA considers the correlation between both tasks, effectively achieving the information interaction for benefiting survival analysis. In short, the contributions of this paper are threefold. (1) We propose a Multimodal Cross-Task Interaction (MCTI) framework that leverages pathological images and genomic data for survival analysis. The framework ingeniously utilizes the subtype classification task to mine valuable disease-positive instances for the survival analysis task, significantly enhancing the model's perception of the tumor microenvironment. (2) We introduce a novel Transport-Guided Attention (TGA) module based on optimal transport theory, highlighting the correlation between subtype classification and survival analysis tasks, effectively performing information interaction for both tasks and optimizing the unified feature representation of pathology and genes. (3) Extensive experiments validate the effectiveness of our approaches, and MCTI outperforms the state-of-the-art frameworks on three public benchmarks. Codes will be publicly available.



\section{Methodology}
\begin{figure}[t]
\includegraphics[width=\textwidth]{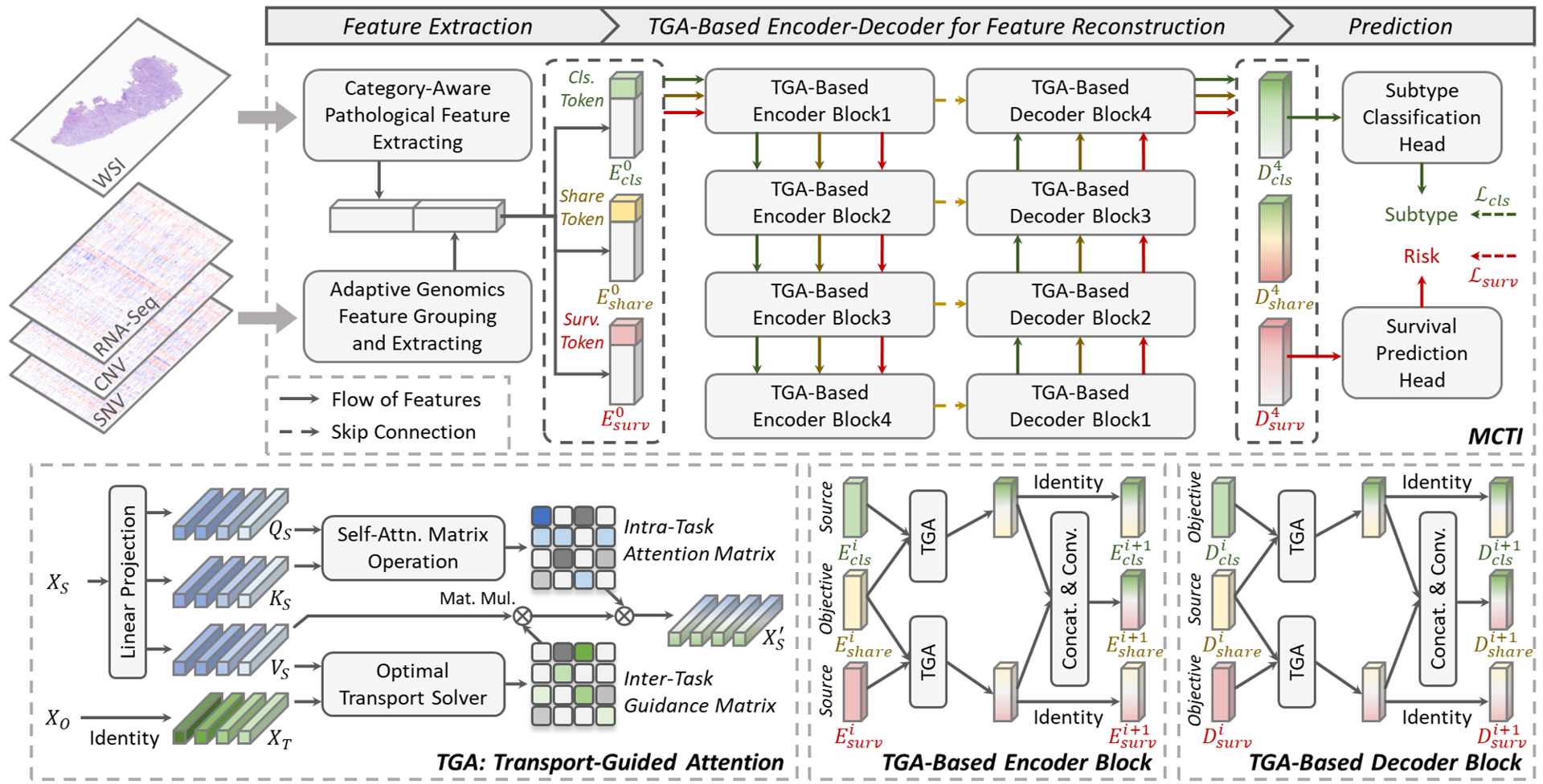}
\caption{
Illustration of Multimodal Cross-Task Interaction (MCTI) framework. ``Mat. Mul.'' denotes matrix multiplication, ``Attn.'' is attention, ``Conv.'' is a convolutional layer, and ``Concat.'' refers to concatenation operation.} 
\label{fig1}
\end{figure}
\subsection{Preliminary}

\subsubsection{Survival Analysis.} Let $D = \left \{ D_{1}, D_{2},..., D_{N}\right \} $ represent the clinical data of $N$ patients. Each patient data can be represented by a $D_{i}=(P_{i}, G_{i},c_{i},t_{i}, Y_{i})$, where $P_{i}$ is the set of WSIs, $G_{i}$ is the set of genomic profiles, $c_{i}\in \left \{ 0,1 \right \} $ is the right uncensorship status, $t_{i}$ is overall survival time, and $Y_{i}$  denotes the subtype of cancer. 
We aim to construct a survival prediction model to estimate $f_{hazard} (T=t|T\ge t, X)$, where $T$ is a random variable and $t$ represents the time point of the occurrence of the death event.
And the cumulative value of the risk function is output as the risk score: $f_{surv} (T\ge t, X)= {\textstyle \prod_{u=1}^{t}} (1-f_{hazard}(T=t|T\ge t, X)).$


\subsubsection{Subtype Classification.} The subtype classification task is often formulated as a MIL problem, where the patches extracted from WSI are considered as instances of the bag, and only the slide-level label $Y$ can be obtained. For a bag $B$ containing $n$ patches, it can be formulated as $B=\left \{ \left ( x_{1}, y_{1} \right ),..., \left ( x_{n}, y_{n} \right ) \right \}, $ where $x_{i}$ represents the $i^{th}$ patch, and $y_{i}$ represents the corresponding label for the patch. 
Mainstream MIL frameworks~\cite{ABMIL,DSMIL} generally adopt a suitable transformation $f$ and a permutation-invariant transformation $g$ to obtain the predicted label $\hat{Y}$ of $B$, given by $\hat{Y} =g\left ( f\left (x_{1}\right ),... f\left (x_{n}\right ) \right ) $.

\subsection{Overall Framework}
The overall framework of our proposed method, Multimodal Cross-Task Interaction (MCTI), is shown in Fig.~\ref{fig1}. Firstly, we extract pathological features using a subtype classification task and obtain genomic features by adaptive grouping. Next, the encoder-decoder structure based on Transport-Guided Attention reconstructs the multimodal embedding. Finally, we use the embedding to predict the survival hazard score. The details are described in the following parts.

\subsection{Pathological and Genomic Feature Extraction}
\label{sec:dsmil}
\subsubsection{WSI Bag Construction.} 
\begin{figure}[t]
\includegraphics[width=\textwidth]{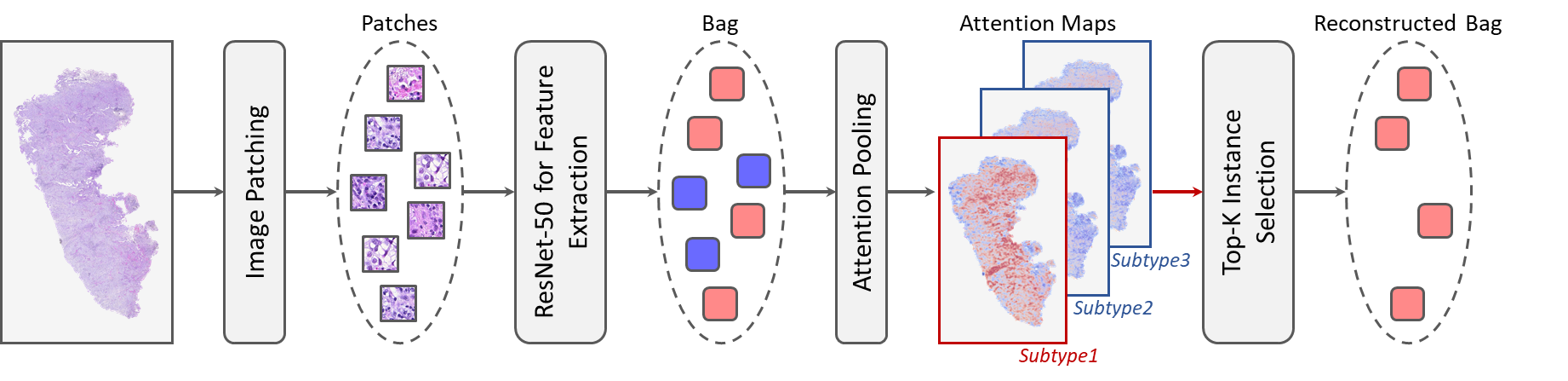}
\caption{
Overall structure diagram of WSI bag construction in MCTI.} \label{fig2}
\end{figure}

We use the DSMIL~\cite{DSMIL} framework as the backbone, where the instance branch assigns an attention score for each instance, reflecting the instance's importance to the subtype classification. We select the top-$k$ instances based on the scores to form a new bag $B_{n}$, as shown in Fig. \ref{fig2}. To supervise the selection of patches related to the tumor microenvironment, we use the cross-entropy $H(y,\hat{y}) = -\sum_{i}y_{i}\cdot log(\hat{y} _{i})$ as loss function for bag classification and instance classification:
$
\mathcal{L}_{DSMIL} = H( Y_{i}, \hat{Y} _{bag}) + H(y_{i}, \hat{y} _{instance}),
\label{eq:dsmil}
$
where $\hat{y} _{instance}$ is the prediction type of the critical instance, $y_{i}$ is equivalent to $Y_{i}$. Then, we can locate relevant regions of the tumor microenvironment of WSI by the subtype classification task. 

\subsubsection{Gene Bag Construction. }

Inspired by recent work~\cite{MCAT,valk2011gene,yao2016imaging}, we aim for our model to adaptively group and extract gene representations. Specifically, we employ Multi-head Self-attention (MSA)~\cite{Transformer} functions in parallel to generate feature representation, which allow the model to consider information from different representation subspaces simultaneously. Then, they are concatenated and once again projected. We obtain the new presentation of genes $B \in\mathbb{ R} ^{ 1\times d} $. 
Here, we replicate the gene embedding $k$ times, resulting in a representation $B \in \mathbb{R}^{k \times d}$ with the same dimensions as the pathological features, where $k$ critical patches have been selected.

\subsection{Multi-Task Encoder-Decoder for Feature Reconstruction}
In the above steps, we derive the representation of the WSI $B\left ( P \right ) = \left \{ p_{1},...,p_{k}  \right \} $ abbreviated as $P$ and genes $B\left ( G \right ) = \left \{ g_{1},...,g_{k}  \right \} $ abbreviated as $G$. Then, we concatenate $B$ and $G$ to obtain a multimodal representation $X \in \mathbb{R}^{ 2k\times d} $. Afterward, we employ a Transport-Guided Attention (TGA) based encoder-decoder for feature reconstruction. It effectively facilitates information interaction to capture task-complementary information and enhance feature representation related to survival analysis.

\subsubsection{Transport-Guide Attention (TGA). }
TGA utilizes optimal transport (OT) to explore the effective knowledge transfer between tasks, yielding cross-task representations. The representations collect complementary information that may not be available within a single task. TGA also employs a self-attention structure to preserve intra-task information within the source distribution, ensuring minimal information loss when it flows from the objective to the source. The computation of the TGA module can be formulated as follows:
\begin{equation}
TGAB\left ( S,O \right ) = \left (  Q^{T}S \right )\left (K^{T}S\right )F_{n}^{T}\left (V^{T}S\right ),
\end{equation}
where $S$ is the source data, $O$ is the objective data. $F_{n}$ is calculated from $W\left ( S,O \right ) = min\left \langle F_{n}, C_{n} \right \rangle_{\mathcal{F} }$, where $< \cdot  >_{\mathcal{F} } $ refers to the Frobenius dot product. The best matching flow $F_{n}$ based on local pairwise similarity and cost matrix $C_{n}$ measures the distance of embeddings using optimal transport theory~\cite{MoCAT,chizat2018scaling}. 

\subsubsection{TGA-based Encoder-Decoder Blocks. }
We hope the encoder can learn the shared inter-task features and the decoder can learn the specific intra-task features layer by layer rather than transferring knowledge between two tasks simultaneously.
Thus, we firstly define three learnable feature tokens  $x_{cls} \in \mathbb{R}^{ k\times d} $, $x_{surv} \in \mathbb{R}^{ k\times d} $, and $x_{share} \in \mathbb{R}^{ k\times d}$ to generate the inputs of encoder:
\begin{equation}
\small
E_{share}^{ 0} =concat(X,x_{share}),E_{cls}^{0} =concat(X,x_{cls}),E_{surv}^{0 } =concat(X,x_{surv}),
\end{equation}
where $E_{share}^{ 0}$ is inter-task embedding, $E_{cls}^{0}$ is the embedding of subtype classification task and $E_{surv}^{0 }$ is the embedding of survival analysis.

The computation of the encoder block is formulated as follows:
\begin{equation}
E_{cls}^{i+1}=TGA\left ( E_{cls}^{i},E_{share}^{i} \right ),
E_{surv}^{i+1}=TGA\left ( E_{surv}^{i},E_{share}^{i} \right ),
\end{equation}
\begin{equation}
E_{share}^{i+1}=conv\left (concat\left (E_{cls}^{i+1},E_{surv}^{i+1}\right) \right).
\end{equation}

The computation of the decoder block is formulated as follows:
\begin{equation}
\small
D_{cls}^{i+1}=TGA(( D_{share}^{i} + E_{share}^{n-i+1}) , D_{cls}^{i}),
D_{surv}^{i+1}=TGA(( D_{share}^{i} + E_{share}^{n-i+1}) , D_{surv}^{i}),
\end{equation}
\begin{equation}
D_{share}^{i+1}=conv\left(concat\left(D_{cls}^{i+1},D_{surv}^{i+1}\right)\right).
\end{equation}
During the decoding process, a skip connection is adopted, and the output of the intermediate step of the encoder is connected to the decoder as part of the input. Our model stacks four layers in the encoder-decoder to obtain more task complementary information.

\subsection{Loss Function}
MCTI uses negative log-likelihood survival loss [9] as the survival analysis loss $\mathcal{L} _{surv}$ and the cross-entropy loss as the loss function of the subtype classification $\mathcal{L} _{cls}$. Combining with $\mathcal{L} _{DSMIL}$ (refer to Section \ref{sec:dsmil}), the total loss $\mathcal{L}_{total}$ can be formulated as:
\begin{equation}
\mathcal{L} _{total}=\mathcal{L} _{cls}+\mathcal{L} _{DSMIL}+\alpha \mathcal{L} _{surv},
\end{equation}
where $\alpha $ is a hyper-parameter for balancing the influence of the loss function.

\section{Experiment}

\subsection{Datasets \& Experimental Settings \& Evaluation Metrics}
We conduct extensive experiments on four public datasets from The Cancer Genome Atlas (TCGA). Specifically, we used Breast Invasive Carcinoma (BRCA), Esophageal Carcinoma (ESCA), Kidney Renal Papillary Cell Carcinoma (KIRP), and Non-small Cell Lung Cancer (NSCLC). We conduct 4-fold cross-validation for evaluation and then randomly split the data as the ratio of training: validation: testing = 60: 15: 25. Our MCTI is implemented in PyTorch 1.12.1 using an NVIDIA RTX 3090 GPU. During training, we use Adam optimization with a learning rate of 0.00005. The number of critical instances selected $k=256$. Following CLAM \cite{CLAM}, we segment tissue regions for each WSI and crop 256 × 256 patches over 20× magnification, then use ImageNet pre-trained ResNet-50 to extract the embedding ($d=1024$) for each patch. 
We use the concordance index (C-Index) to evaluate the performance of survival analysis models. Kaplan-Meier analysis is utilized to measure the statistical significance between low risk group and high risk group.

\subsection{Comparison with State-of-the-Art Methods}
Table \ref{tab:table1} shows the quantitative results of unimodal and multimodal methods. For unimodal methods, we adopt SNN \cite{SNN} as the genomic features extractor for survival analysis. Additionally, AvgPool, ABMIL~\cite{ABMIL}, CLAM~\cite{CLAM}, and DSMIL~\cite{DSMIL} use pathological images for survival analysis. Table \ref{tab:table1} shows that their performances are worse than multimodal methods.

We also compare our method with multimodal survival analysis frameworks, including Porpoise \cite{Porpoise}, MCAT \cite{MCAT}, M3IF \cite{M3IF}, CMAT \cite{CMTA}. 

\begin{figure}[t]
\includegraphics[width=\textwidth]{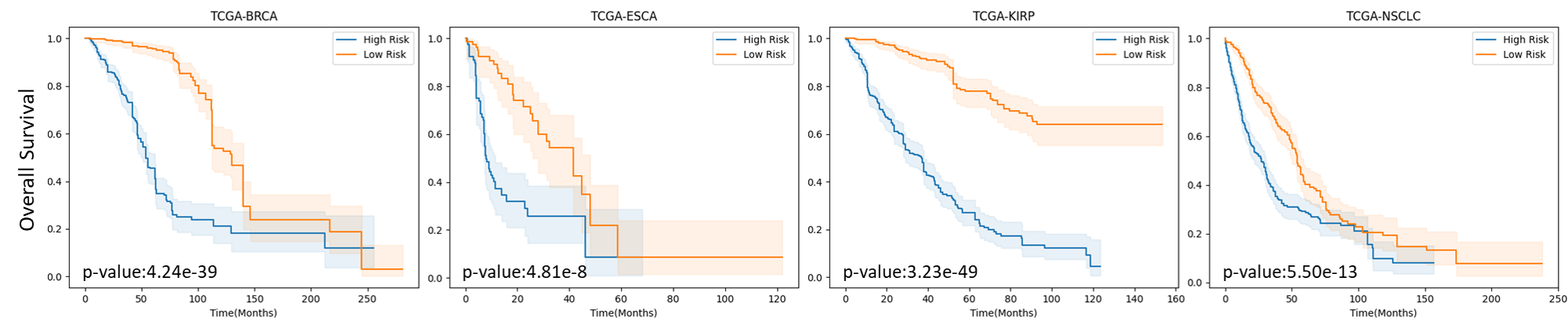}
\caption{Kaplan-Meier Analysis on four cancer datasets according to predicted risk scores. P-value < 0.05 means significant statistical difference between low-risk (blue) and high-risk (red).} 
\label{fig3}
\end{figure}

Unlike them, exploring modality alignment to transfer potential complementary information, our method introduces the subtype classification task to assist survival analysis. Our model outperforms other state-of-the-art multimodal methods by a large margin in the BRCA, ESCA, and KIRP datasets. Especially on the BRCA dataset, our model surpasses the second best by 7.7\%. We also visualize the Kaplan-Meier survival curves in Fig. \ref{fig3} to demonstrate a statistical distinction in patient stratification performance. 

\begin{table}[t!]
  \centering
  \caption{C-Index (mean $\pm$ std) performance over four cancer datasets. The best results are shown in \textbf{bold}, and the second best ones are \underline{underlined}.}
  \label{tab:table1}
  \resizebox{1\textwidth}{!}{
    \begin{tabular}{l|c|c|c|c} 
      \specialrule{.1em}{.05em}{.05em}
      \textbf{Model} & \textbf{TCGA-BRCA} & \textbf{TCGA-ESCA} & \textbf{TCGA-NSCLC} & \textbf{TCGA-KICA}\\
      \specialrule{.1em}{.05em}{.05em}
      SNN & $0.584\pm0.035$ & $0.540\pm0.056$ & $0.542\pm0.026$ & $0.651\pm0.011$\\
      \specialrule{.1em}{.05em}{.05em}
      AvgPool & $0.566\pm0.083$ & $0.567\pm0.024$ & $0.568\pm0.028$ & $0.658\pm0.049$\\
      AttnMIL~\cite{ABMIL} & $0.556\pm0.100$ & $0.583\pm0.036$ & $0.571\pm0.030$ & $0.657\pm0.073$ \\
      CLAM-SB~\cite{CLAM} & $0.489\pm0.083$ & $0.551\pm0.051$ & $0.552\pm0.037$ & $0.692\pm0.058$\\
      CLAM-MB~\cite{CLAM} & $0.563\pm0.066$ & $0.591\pm0.041$ & $0.581\pm0.025$ & $0.695\pm0.040$\\
      DSMIL~\cite{DSMIL} & $0.543\pm0.095$ & $\underline{0.594}\pm0.007$ & $0.581\pm0.027$ & $0.645\pm0.033$\\
      \specialrule{.1em}{.05em}{.05em}
      MCAT~\cite{MCAT} & $0.544\pm0.034$ & $0.553\pm0.085$ & $0.638\pm0.013$ & $0.693\pm0.010$\\
      CMTA~\cite{CMTA} & $0.578\pm0.029$ & $0.565\pm0.033$ & $\textbf{0.655}\pm0.028$ & $\underline{0.698}\pm0.015$\\
      PORPOISE~\cite{Porpoise} & $0.587\pm0.043$ & $0.527\pm0.034$ & $0.542\pm0.013$ & $0.677\pm0.040$\\
      M3IF~\cite{M3IF} & $\underline{0.579}\pm0.047$ & $0.557\pm0.031$ & $0.502\pm0.027$ & $0.664\pm0.018$\\
      \specialrule{.1em}{.05em}{.05em}
      Ours & $\textbf{0.656}\pm0.018$ & $\textbf{0.621}\pm0.044$ & $\underline{0.639}\pm0.028$ & $\textbf{0.723}\pm0.046$\\
      \specialrule{.1em}{.05em}{.05em}
    \end{tabular}
  }
\end{table}

\subsection{Ablation Studies}
In this section, we perform ablation experiments to validate the impact of our modules. Quantitative results are presented in Table \ref{tab:table2}.

\begin{figure}[t]
\centering
\resizebox{0.8\textwidth}{!}{\includegraphics[width=\textwidth]{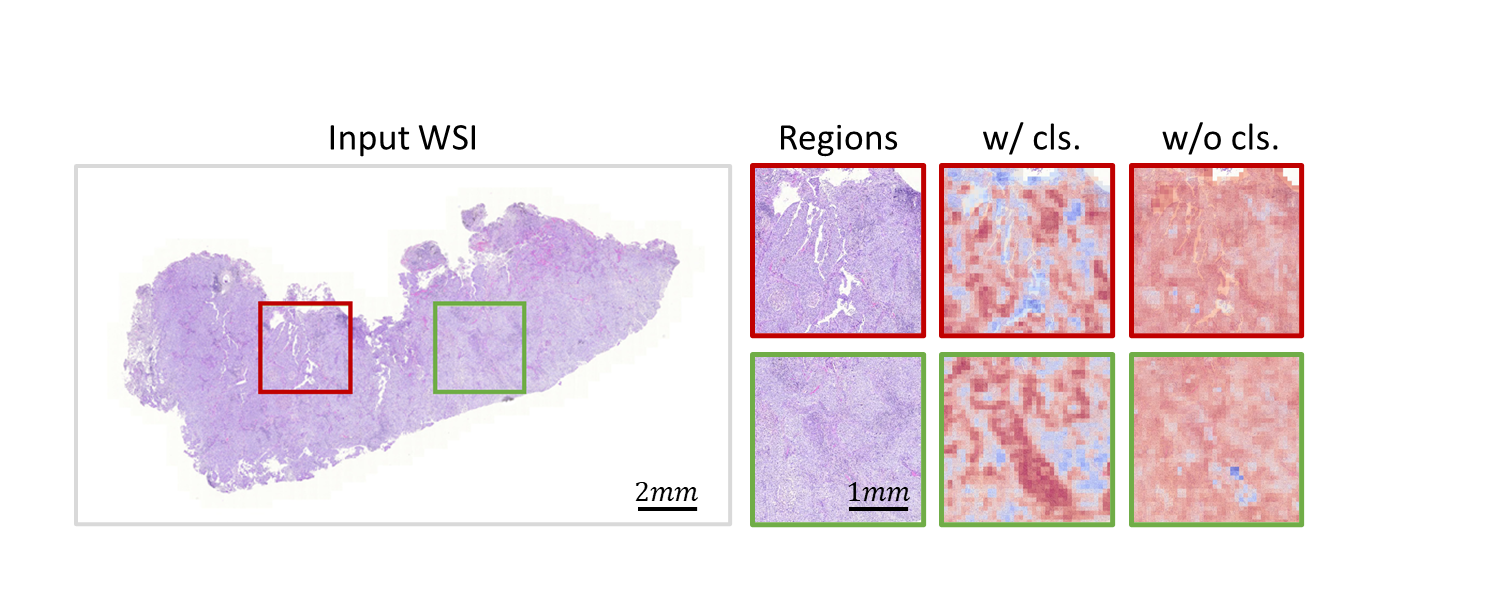}}
\caption{Visualization of attention map. ``cls.'' is the subtype classification task.
} 
\label{fig4}
\end{figure}

\begin{table}[t]
  \centering
  \caption{Ablation study assessing C-Index (mean ± std) performance. The best results are shown in \textbf{bold}. ``recon.'' refers to reconstruction.}
  \label{tab:table2}
  \resizebox{1\textwidth}{!}{
  \begin{tabular}{l|c|c|c|c} 
  \specialrule{.1em}{.05em}{.05em}
    \textbf{Settings} & \textbf{TCGA-BRCA} & \textbf{TCGA-ESCA} & \textbf{TCGA-KICA} & \textbf{TCGA-NSCLC}\\
    \specialrule{.1em}{.05em}{.05em}
    MCTI w/o WSI Recon. & $0.489\pm0.083$ & $0.551\pm0.051$ & $0.692\pm0.058$ & $0.552\pm0.037$\\
    MCTI w/o Genes Group. & $0.563\pm0.066$ & $0.591\pm0.041$ & $0.695\pm0.040$ & $0.581\pm0.025$\\
    MCTI w/o TGA & $0.545\pm0.021$ & $0.592\pm0.093$ & $0.582\pm0.148$ & $0.537\pm0.020$\\
    MCTI & $ \textbf{0.656}\pm0.018$ & $ \textbf{0.621}\pm0.044$ & $\textbf{0.723}\pm0.046$ & $\textbf{0.639}\pm0.028$\\
    \specialrule{.1em}{.05em}{.05em}
  \end{tabular}}
\end{table}

\textbf{Effectiveness of Critical Patch Selection.}
Table \ref{tab:table2} shows that when patches are randomly selected, C-Index drops by 16.7\%, 7.0\%, 3.1\%, and 8.7\% in the four datasets. In contrast, choosing critical patches based on subtype classification aims to explore the tumor microenvironment and provide a better WSI bag representation for survival analysis tasks.

\textbf{Effectiveness of Multi-Head Attention on Genomic Embedding.}
As shown in Table \ref{tab:table2}, the BRCA dataset is significantly influenced by multi-head attention. Furthermore, the figure indicates that adaptive genes grouping and feature extraction are beneficial for survival analysis.

\textbf{Effectiveness of Subtype Classification in TGA.}
When we remove the loss function of the arbitrary task in TGA, for the sake of fairness of the model, we also replace the branch of the classification task with survival analysis, i.e., $\mathcal{L} _{total}=\mathcal{L} _{surv}+\alpha \mathcal{L} _{surv}$. 
From Table \ref{tab:table2}, we can find that the performance decreases by 11.1\%, 2.9\%, 14.1\%, and 10.2\% on the four datasets respectively.
Fig. \ref{fig4} shows the attention map generated by DSMIL, from which we can observe that our model excels in capturing tumor-related regions.

\section{Conclusion}

In this study, we propose a novel Multimodal Cross-Task Interaction (MCTI) framework that leverages subtype classification as an auxiliary task to enhance survival analysis.
Based on attention-based multiple instance learning, MCTI performs subtype classification to precisely identify tumor regions within WSIs, enhancing the representation of TME-related features.
Furthermore, a Transport-Guided Attention (TGA) module is designed to consider the correlation between tasks and effectively transfer the knowledge from the subtype classification task to survival analysis.
Our experiments demonstrate the effectiveness of MCTI, outperforming state-of-the-art frameworks across three public benchmarks. This study provides fresh insight into survival analysis. Future work would focus on explaining the relations between subtype classification and survival analysis and validating MCTI's performance on multi-center datasets.

\bibliography{r1}

\end{document}